\documentclass{hotnets22}
\pdfoutput=1
\usepackage{times}
\usepackage{hyperref}
\usepackage{outlines}
\usepackage{titlesec}

\usepackage[dvipsnames]{xcolor}
\usepackage{xspace}
\usepackage{outlines}
\usepackage{enumitem} 
\usepackage{subcaption}
\usepackage{algorithm}
\usepackage{siunitx}
\usepackage[noend]{algpseudocode}
\usepackage{caption}
\usepackage{paralist}
\usepackage{url}

\usepackage{wrapfig}

\definecolor{DrawBlue}{HTML}{DAE8FC}
\definecolor{DrawRed}{HTML}{F8CECC}
\definecolor{DrawPurple}{HTML}{E1D5E7}

\newcommand{\notes}[1]{}
\renewcommand{\notes}[1]{#1} 

\newcommand{\ccfuzz}[0]{CC-Fuzz\xspace}

\newcommand{\cut}[1]{} 

\newcommand\vertarrowbox[3][6ex]{%
  \begin{array}[t]{@{}c@{}} #2 \\
  \left\uparrow\vcenter{\hrule height #1}\right.\kern-\nulldelimiterspace\\
  \makebox[0pt]{\scriptsize#3}
  \end{array}%
}

\hypersetup{draft}

\begin{document}
\pagenumbering{arabic}


\title{\ccfuzz: Genetic algorithm-based fuzzing for stress testing congestion control algorithms.}

\author{
  Devdeep Ray\\
  Carnegie Mellon University\\
  \texttt{devdeepr@cs.cmu.edu}
  \and
  Srinivasan Seshan\\
  Carnegie Mellon University\\
  \texttt{srini@cs.cmu.edu}
}

\maketitle
\begin{abstract}
    Congestion control research has experienced a significant increase in interest in the past few years, with many purpose-built algorithms being designed with the needs of specific applications in mind. These algorithms undergo limited testing before being deployed on the Internet, where they interact with other congestion control algorithms and run across a variety of network conditions.
This often results in unforeseen performance issues in the wild due to algorithmic inadequacies or implementation bugs, and these issues are often hard to identify since packet traces are not available. 

In this paper, we present \ccfuzz, an automated congestion control testing framework that uses a genetic search algorithm in order to stress test congestion control algorithms by generating adversarial network traces and traffic patterns.
Initial results using this approach are promising - \ccfuzz automatically found a bug in BBR that causes it to stall permanently, and is able to automatically discover the well-known low-rate TCP attack, among other things.
\end{abstract}
\section{Introduction}
\label{sec:introduction}


Traditional congestion control algorithms (CCAs) were designed with the core goals of high throughput and fairness, while preventing congestion collapse.
Unfortunately, traditional loss-based CCAs fall short in terms of meeting the performance requirements for many emerging applications and network environments.
Recent research has shown a significant interest in designing new congestion control algorithms for achieving specific performance goals, or improving general CCA performance.
For example, SCReAM~\cite{johansson2017self}, GoogCC~\cite{carlucci2016analysis}, and Sprout~\cite{winstein2013stochastic} are designed for low latency video streaming, with the key goal of achieving low end-to-end delay.
Some CCAs are designed to extract maximum performance from special networking infrastructure avaiable in data center settings, such as programmable NICs (Swift, TIMELY), switches supporting AQM (DCTCP), and hybrid optical-packet networks~\cite{mukerjee2020adapting}.
Other CCAs like Copa~\cite{arun2018copa}, Nimbus~\cite{goyal2018elasticity}, and TCP-BBR~\cite{cardwell2017bbr} use complex network modeling techniques in order to achieve dual goals of (1) low delay, and (2) high throughput when competing with other flows.

A significant fraction of new CCAs are developed by the academic community, where the opportunities for large scale testing in the real-world are limited.
In order to deploy a new CCA on the Internet or in large-scale data centers, where the packets traverse across a variety of network conditions and encounter diverse cross traffic patterns, it is important to evaluate the robustness of a CCA and it's implementation across a wide range of scenarios.
This is a challenging task in an academic setting - many newly proposed CCAs are evaluated using a combination of small scale deployment~\cite{yan2018pantheon}, and local scenario-based emulation and simulation~\cite{netravali2014mahimahi}.
These new CCAs are much more complex compared to traditional loss-based CCAs like TCP-CUBIC and TCP-Reno, and the tests typically performed at an academic scale can easily miss situations where the algorithm fails to achieve it's goals (like high utilization, fairness, or low delay~\cite{ware2019modeling, goyal2018elasticity}), or corner cases where bugs in the CCA implementation are triggered. 

In this paper, we describe the design of our testing framework called ``\ccfuzz\footnote{\ccfuzz is a pun on Sisyphus. Wikipedia notes, ``tasks that are both laborious and futile are therefore described as Sisyphean''~\cite{enwiki-sisyphus}}'', and demonstrate the value of using genetic search algorithms for exploring the search space of link behavior and cross traffic patterns in order to identify issues with CCAs and their implementations, or inspire confidence in a CCA before it is deployed in the real-world.
A genetic algorithm is a search heuristic that is inspired from the Darwinian theory of biological evaluation - the algorithm maintains a pool of traces and on every iteration, each entity in the population is assigned a fitness score. The fitness scores are used to generate the next population generation in a manner that is similar to natural selection.
In our case, the population entities are network traces, the fitness scores are based on the performance of a CCA for each trace, and evolution involves modifying the traces in the population in a manner such that eventually we find traces that trigger poor behavior in the CCA being tested (convergence).

\ccfuzz has two modes - (1) Link mode aims to identify bottleneck packet transmission patterns, and (2) Traffic mode aims to identify cross traffic patterns, that that result in poor performance for the particular CCA being tested.
We believe these two approaches can trigger different behaviors, since variations in the link rate model arbitrary delay jitter, whereas the delay is bounded when injecting cross traffic.
In order to generate realistic link behavior and cross traffic, \ccfuzz uses (1) heuristics during trace generation, and (2) leverages the generality of genetic algorithms by using carefully designed fitness scores for imposing implicit constraints that are hard to model using heuristics. In Section~\ref{sec:discussion}, we propose an alternate way to impose realism on network traces as part of future work.


The current version of \ccfuzz uses NS3-based~\cite{carneiro2010ns} simulation in order to evaluate CCAs and assign fitness scores for the traces, and we evaluate the pre-defined CCAs in NS3.
In the future, we plan to use emulation-based testing of CCA implementations, since the versions of CCAs in NS3 can sometimes differ from their real implementations.
Our findings (\S~\ref{sec:results}) include:
\begin{compactenum}
    \item \textbf{BBR}: \ccfuzz is able to generate network traces that cause BBR to permanently stall due to the way ACKs and spurious retransmissions interact with each other during a retransmission timeout. \ccfuzz is also able to generate traffic patterns that trigger BBR to cause high queuing delays.
    \item \textbf{CUBIC}: \ccfuzz is able to generate traffic patterns that trigger a NS3-specific implementation bug in CUBIC regarding CWND updates.
    \item \textbf{Reno}: \ccfuzz is able to generate traffic patterns that are similar to the TCP low-rate attack~\cite{kuzmanovic2003low}.
\end{compactenum}
In the remainder of the paper, we discuss the design of \ccfuzz (\S~\ref{sec:design}) and present directions for future work (\S~\ref{sec:discussion}).
\section{Motivation and Related Work}
\label{sec:background}
Newly proposed CCAs are often evaluated using metrics such as throughput, delay and fairness across a range of simple scenarios like testing a CCAs ability to track available bandwidth at macroscopic time-scales, and coexist with other flows (e.g. by introducing competing flows using various CCAs).


Past work has shown that commonly evaluated scenarios often fail to catch surprising failure modes - In \cite{ware2019modeling}, the authors use mathematical modeling to show that multiple BBR flows are unfair towards loss-based CCAs.
In CCAC~\cite{ccac}, the authors argue that basic evaluation techniques are not sufficient for capturing every scenario that causes undesirable behavior, and propose a formal verification technique that can generate network behavior that satisfy queries about CCA performance.
Formal approaches are limited since they analyze ``theoretical models'' of CCAs.
This step can hide key bugs and issues present in real implementations.
In addition, formal techniques become intractable when the fidelity of the model is increased or when verifying properties over long time periods (e.g. BBR's minRTT probing behavior).

Fuzzing~\cite{sutton2007fuzzing} is a widely used technique for discovering vulnerabilities in code.
TCPwn~\cite{jero2018automated} uses model-based fuzzing in order to identify manipulation attacks (e.g. dup ACK injection, ACK storm, sequence desynchronization) on CCAs.
TCP-Fuzz~\cite{zou2021tcp} tests TCP stack implementations for bugs, and ACT~\cite{sun2019model} uses state space exploration to generate particular numerical values of the state variables in the CCA implementation that triggers buggy behavior.

Packetdrill~\cite{cardwell2013packetdrill} uses scripted tests to detect bugs in the networking stack and for regression testing of CCAs.
Packetdrill has proven successful in catching many issues over time, but it requires clearly laid out networking scenarios that must be developed by hand - this can miss many situations.

The goal of \ccfuzz is to find realistic situations where CCA performance suffers due to packet delivery timing and losses \textit{automatically}.
We do not aim to find protocol-level bugs that are triggered by injecting spoofed packets - the tools mentioned above can be used for low-level bug finding.
We are not aware of any existing system that automatically generates network environments for stress-testing CCAs for high level throughput and other performance objectives.
\section{Design}
\label{sec:design}
\begin{figure}
\begin{algorithm}[H]
\renewcommand{\thealgorithm}{}
    \caption{Genetic Algorithm Loop}
    \begin{algorithmic}
            \Procedure{\ccfuzz}{}
            \State $\textsc{traces} \gets \text{Initial pool of traces}$
            \State $\textit{kElite} \gets \text{Number of traces that live on unmodified.}$
            \State $\textit{kCrossover} \gets \text{Count of new traces generated by}$\newline
        \hspace*{7em}$\text{combining traces with high scores.}$
            \Repeat \For {$\textit{trace} \in  \textsc{traces}$} \State {
                $\textsc{score}(i) \gets \text{Score when CCA run  with }$\newline
        \hspace*{9em}
                $\textsc{traces}(i)$
            } 
            \EndFor
            \State $\textsc{elite} \gets \text{Top \textit{kElite} traces}$
            \State $\textsc{crossover} \gets \textit{kCrossover} \text{ traces that are}$\newline
            \hspace*{8em} generated by combining traces
            \State $\textsc{mutated} \gets \textit{len(\textsc{traces}) - kElite - kCrossover}$\newline
            \hspace*{7em} $\text{traces generated by modifiying traces}$
            \State $\textsc{traces} \gets \textsc{elite} + \textsc{crossover} + \textsc{mutated}$
            \Until{convergence}
            \EndProcedure
    \end{algorithmic}
\end{algorithm}
\vspace{-0.1in}
\caption{}\label{design:ga-loop}
\end{figure}


\ccfuzz explores the search space of network behavior and cross traffic by using a genetic algorithm for generating network traces in order to identifying network behaviors that cause the CCA to perform poorly.
\ccfuzz's high level loop is described in Figure~\ref{design:ga-loop}. 
\ccfuzz's core components include the following: 
\begin{compactenum}
    \item \textbf{Trace Generator}: Generates initial traces, and defines functions for performing cross-overs between pairs of traces, and mutating individual traces.
    \item \textbf{Scoring Function}: Simulates or emulates the CCA's performance for a link or traffic trace, and assigns a score based on the property being evaluated (e.g. throughput, delay, loss, or a combination).
    \item \textbf{Selection Algorithm}: Selects trace pairs for generating cross-over traces for the next generation, and selects traces that will be mutated before being added to the next generation trace pool.
\end{compactenum}
These components are discussed in further detail below.


\subsection{Network Model}
\ccfuzz uses a simple network topology with two sources (one source uses the CCA being tested, and the other source generates cross traffic) that are connected to a gateway with high speed links.
The gateway is connected to a sink via a bottleneck link with a fixed propagation delay.
The gateway consists of a fixed-size drop-tail FIFO queue.

\ccfuzz has two distinct modes - 
\begin{compactenum}
\item \textbf{Link Fuzzing: } We generate bottleneck service curves that define the rate at which packets are drained from the bottleneck queue and transmitted over the link.
\item \textbf{Traffic Fuzzing: } We generate traffic traces that determine the injection of cross traffic into the bottleneck queue, and the bottleneck transmission rate is fixed.
\end{compactenum}
We explore two different modes due to the following reasons:
\begin{compactenum}
    \item A fixed rate link with variable cross traffic cannot model unbounded delays that can be caused by variable links with a fixed bottleneck buffer size.
    \item Variable links with a fixed bottleneck buffer size cannot model loss with bounded delay, which can be caused by queue-building cross traffic.
    \item A realistic link may exhibit aggregation and delay jitter, and some degree of long-term temporal rate variation. On the other hand, realistic cross-traffic can be highly adversarial. Separating out link and traffic fuzzing enables us to model such constraints and makes the results easier to understand.
\end{compactenum}


These two modes are quite general, but they do not capture certain behaviors like random packet losses. We defer this and other approaches like combining link and traffic fuzzing to future work (\S~\ref{sec:discussion}).


\subsection{Link Fuzzing}
Link fuzzing aims to generate bottleneck service curves that trigger poor performance in the CCA being tested. We represent the service curve as a sequence of packet transmissions (similar to the model used in MahiMahi~\cite{netravali2014mahimahi}).
This representation lends itself well for modeling jitter, and imposing high-level constraints on the service curve.
For link fuzzing, we fix the total number of packets that can be serviced by the link during a run (and thus, the average bandwidth).

\begin{figure}
\begin{algorithm}[H]
\renewcommand{\thealgorithm}{}
    \caption{Packet Distribution Algorithm}
    \begin{algorithmic}
            \Procedure{DistPackets}{$\textit{num, start, end}$}
            \If {$\textit{num} == 0$}
                \Return $[\;]$
            \EndIf
            \If {$\textit{num} == 1$}
                \Return $[\frac{\textit{start}+\textit{end}}{2}]$
            \EndIf
            \State $\textit{rate} \gets \frac{\textit{num}}{\textit{end} - \textit{start}}$
            \Loop
                \State $\textit{tsplit} \gets \textbf{U}(\textit{start}, \textit{end})$ 
                \State $\textit{numleft} \gets \textbf{U}(0, \textit{num})$ 
                \State\Comment{\textbf{U} is uniform random sampling.}
                \If{$\textit{end} - \textit{start} < \textit{kAgg}$}
                    \textbf{break}
                \EndIf
                \State $\textit{lrate}\gets\frac{\textit{numleft}}{\textit{tsplit}-\textit{start}} , \textit{rrate}\gets\frac{\textit{num - numleft}}{\textit{end}-\textit{tsplit}}$
                \vspace{0.05in}
                \If{$ \textit{lrate}>2\times\textit{rate} \textbf{ or } \textit{rrate}>2\times\textit{rate}$}
                    \State{\textbf{continue}}
                \EndIf
                \If{$ \textit{lrate}<0.5\times\textit{rate} \textbf{ or } \textit{rrate}<0.5\times\textit{rate}$}
                    \State{\textbf{continue}}
                \EndIf
            \EndLoop
            \State \Return \textsc{DistPackets}(\textit{numleft}, \textit{start}, \textit{tsplit})\newline
        \hspace*{5em} + \textsc{DistPackets}(\textit{num - numleft}, \textit{tsplit}, \textit{end})
            \EndProcedure
    \end{algorithmic}
\end{algorithm}
\vspace{-0.1in}
\caption{}\label{design:dist-packets-loop}
\end{figure}

\begin{figure}
    \centering
    \vspace{0.07in}
    \subcaptionbox{5 second interval.\label{fig:design:5sdistpackets}}{\includegraphics[width=0.48\columnwidth]{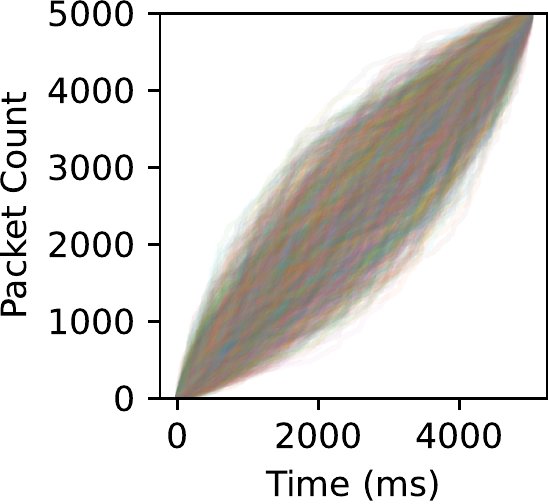}}%
    \hfill\subcaptionbox{50 millisecond interval.\label{fig:design:50msdistpackets}}{\includegraphics[width=0.445\columnwidth]{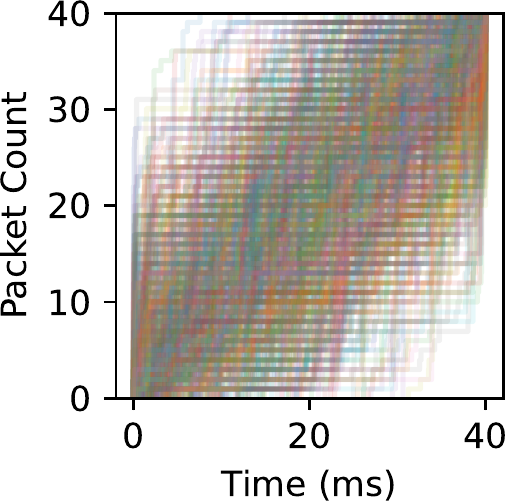}}
    \caption{Service curves generated using \textsc{DistPackets}, with an average rate of 12 Mbps and $\textit{kAgg}=\SI{50}{\milli\second}$.} 
    \label{fig:design:service-curves}
\end{figure}

\textbf{Initial Trace Generation.}
In order to generate realistic traces, but still cover a large portion of the search space, \ccfuzz distributes the packet transmissions over time using the \textsc{DistPackets} algorithm shown in Figure~\ref{design:dist-packets-loop}.
The key idea is to recursively divide packets by splitting the time length and number of packets into two in each step, and ensuring that in each step, the average rate for each division lies within a multiplicative range of the average rate.
This mechanism is a heuristic that bounds the long term variation in the bandwidth.
Deeper in the recursion, when the time length drops below a threshold (\textit{kAgg}), the bound checks are relaxed in order to allow arbitrary short-term rate variations to model packet aggregation and jitter.
In Figure~\ref{fig:design:service-curves}, we show the distribution of service curves generated by this algorithm.

\textbf{Evolution Mechanism}
Typical genetic algorithms have two evolution mechanisms: mutations, and crossovers.
Mutation mechanisms choose entities that have the desired properties, and modify them in order to generate new entities for populating the next generation.
Crossover mechanisms pick two or more traces that have the desired properties, and combine them in some way for generating new entities.

When creating a new generation from a pool of link traces from a previous generation, \ccfuzz must ensure that the same properties as that of the initial generation hold - otherwise, the constraints we want to impose on the traces can be violated to arbitrary extents after only a few generations.
For mutating a link trace, \ccfuzz selects a random split point in the trace, and redistributes packets (\textsc{DistPackets}) on either the left or the right side of the split point (chosen using a coin toss).
This preserves the properties of the trace from the initial generation step.
\ccfuzz does not use crossovers for link fuzzing, since there is no obvious way of combining two independent service curves while maintaining the core properties of the two subtraces(e.g. total packets transmitted, rate variation heuristics).
In order to generate easier to understand link traces, \ccfuzz optionally supports trace annealing.
After evaluating a trace and before performing mutations, \ccfuzz applies Gaussian smoothing to the packet timestamps.
Over multiple generations, this reduces the link variation in regions that are not relevant for triggering the poor behavior.

\subsection{Traffic Fuzzing}
\ccfuzz uses the same algorithm (\textsc{DistPackets}) for generating traffic traces, with some modifications. 
\begin{compactenum}
    \item \textbf{Trace generation heuristics: } We eliminate the local rate constraints, allowing bursty cross traffic.
    \item \textbf{Crossover operation: } 
    By eliminating local rate constraints, we define the crossover operation as follows: randomly choose a split point by packet count, randomly select the left half of one trace and the right half of the other trace around the split point, and combine the two sets of timestamps.
\end{compactenum}

In the case of traffic fuzzing, it is desirable to generate ``minimal'' traffic injection vectors that induce poor behavior in CCAs.
For example, a large burst of cross traffic where many packets of the cross traffic are lost will have the same impact if the lost packets were never sent.
In addition, cross traffic arrives and departs the bottleneck queue when the CCA under test is silent (e.g. when TCP is waiting for ACKs after filling the CWND) has no impact.

In order to impose these properties, we allow a variable number of cross traffic packets up to a maximum limit.
When regenerating a portion of the trace during a mutation operation, the number of packets in that portion are changed randomly.
During a crossover operation, the number of traffic packets naturally change based on whether the trace on the right side has more or less packets.
This is combined with a change to the scoring function (\S~\ref{sec:design:scoring-function}) in order to implicitly impose constraints like minimizing the number of cross-traffic packets required to trigger poor performance in the CCA being tested.


\subsection{Scoring Function}
\label{sec:design:scoring-function}
The scoring function is a key aspect of a genetic algorithm - it determines which traces were successful in triggering specific performance behavior, and allows implicit modeling of desirable properties in a link or traffic trace.
As part of calculating the score for a trace, \ccfuzz runs the CCA using the link or traffic trace (simulated using NS3. Emulation using tools like MahiMahi can also be used - comparison in Section~\ref{sec:design:emulation-vs-simulation}), and analyzes the queuing behavior.
The score assigned to each trace in a generation has two components: performance score and trace score. 

\textbf{Performance Score.}
The performance score can be designed for specific types of poor behavior like high loss rate, high delay or low utilization.
For example, for quantifying low utilization, \ccfuzz calculates windowed throughput for the run, and takes the average of the lowest 20\% of the windows. 
Compared to using the overall throughput, this prevents the algorithm preferring traces that trigger poor behavior early on, improving trace diversity.

\textbf{Trace Score.}
In order to model properties of the traces that are hard to model during trace generation, each trace can be assigned a score based on how well it satisfies the desired properties.
For example, \ccfuzz scores traffic traces using the (negation of) total traffic packets and the total traffic packets dropped in order to make the genetic algorithm gravitate towards generating minimal traffic vectors.

\subsection{Selection Algorithm}
Once the traces in a generation have been assigned scores, we rank the traces from highest score to lowest score. 
We first pick \textit{kElite} of the highest scored traces that make it to the next generation unchanged. 
We assign a relative probability of $\frac{1}{\textit{rank}}$ to each trace and then choose \textit{kCrossover} pairs of traces according to these probabilities, and combine them for generating crossover traces.
The same probabilities (based on rank) are used for picking traces that undergo mutation for generating the rest of the traces in order to maintain a constant population size.

\subsection{Emulation vs. Simulation}
\label{sec:design:emulation-vs-simulation}
Our current implementation of \ccfuzz uses NS3-based simulation for evaluating a CCA's performance for a link or traffic trace.
An alternative is to use a network emulator like MahiMahi.
In either case, \ccfuzz will test a combination of the CCA implementation and the run-time framework, finding failures in either system and their interactions.

The benefit of emulation is the ability to test a real implementation of a CCA.
Unfortunately, emulating multiple traces in parallel in a reproducible manner is challenging. We need to ensure that the performance is not affected due to CPU and memory bottlenecks, and that the start time of a flow is synchronized with the network trace.
Otherwise, the CCA behavior can be very different across generations for a given trace, which can delay or even prevent convergence of the genetic algorithm.

Simulation, on the other hand, will generate identical results across repeated runs, resulting in faster convergence.
In addition, for link rates in 10s of Mbps, simulation is likely to be faster than real-time emulation, and the results of the simulation are not affected by machine load - this makes it easy to massively parallelize the algorithm on a single machine.
The key drawback of simulation is that it does not test the actual implementation, but a re-implementation in the simulation framework (e.g. NS3).
Tools like DCE~\cite{tazaki2013dce} can mitigate this drawback by simulating real network stacks.

In addition, randomization in a CCA's implementation can also prevent convergence.
In such cases, we need to modify the CCA implementation so that the randomization is repeatable (fix the random seed).
This is much easier in a simulated setup as opposed to modifying kernel CCA code in an emulated environment.
In the future, we plan to explore the use of emulation for \ccfuzz.

\section{Findings}
\label{sec:results}
In this section, we will discuss some interesting findings that \ccfuzz was able to automatically discover.
For all of our tests, we set the bottleneck bandwidth to 12 Mbps (average bandwidth in the case of link fuzzing) and set the propagation delay of the bottleneck link to 20 ms.
TCP-SACK and delayed ACKs are enabled (Linux defaults), and min-RTO is set to 1 second (as per RFC 6298/2.4, Linux uses 200 ms). We use a population size of 500, and use an island-isolation~\cite{whitley1999island} strategy with 20 islands for solution diversity, where 10\% of the traces migrate every 10 generations. Across island generations, the best trace is preserved (\textit{kElite = 1}), 30\% of the traces are crossovers, and the rest are mutations.

\subsection{BBR - Stuck Throughput}
\begin{figure*}
    \centering
    \subcaptionbox{\ccfuzz traffic trace that causes BBR to get stuck.\label{fig:bbr-traffic}}{\includegraphics[width=0.38\columnwidth]{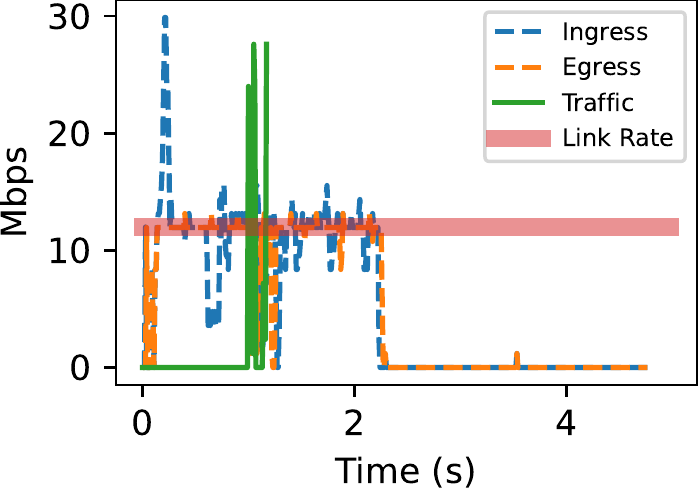}}\hfill%
    \subcaptionbox{\ccfuzz link trace that causes BBR to get stuck.\label{fig:bbr-link}}{\includegraphics[width=0.38\columnwidth]{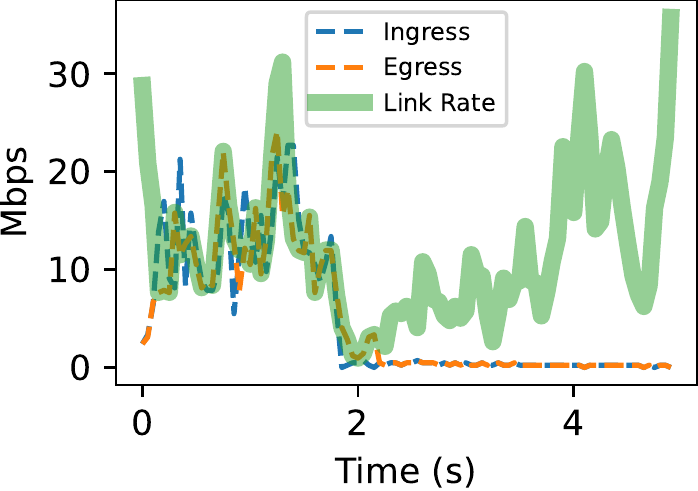}}\hfill%
    \subcaptionbox{Timeline showing how BBR's bug is triggered.\label{fig:bbr-timeline}}{\includegraphics[width=0.38\columnwidth]{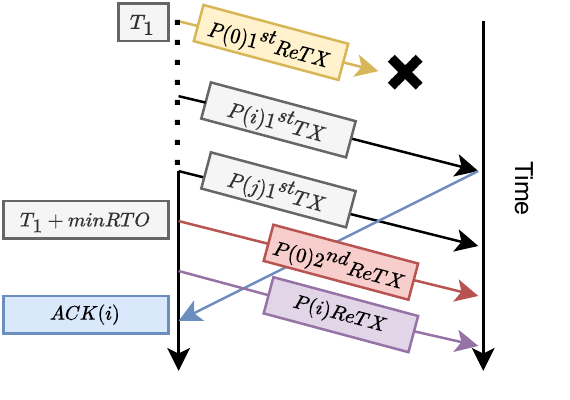}}\hfill%
    \subcaptionbox{\ccfuzz performance with and without BBR patch.\label{fig:bbr-patch}}{\includegraphics[width=0.42\columnwidth]{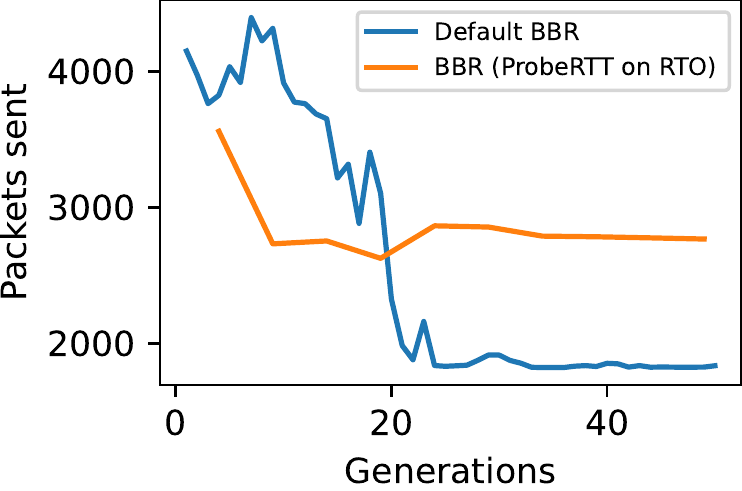}}\hfill%
    \subcaptionbox{\ccfuzz triggering high delays in BBR with cross traffic.\label{fig:bbr-delay}}{\includegraphics[width=0.4\columnwidth]{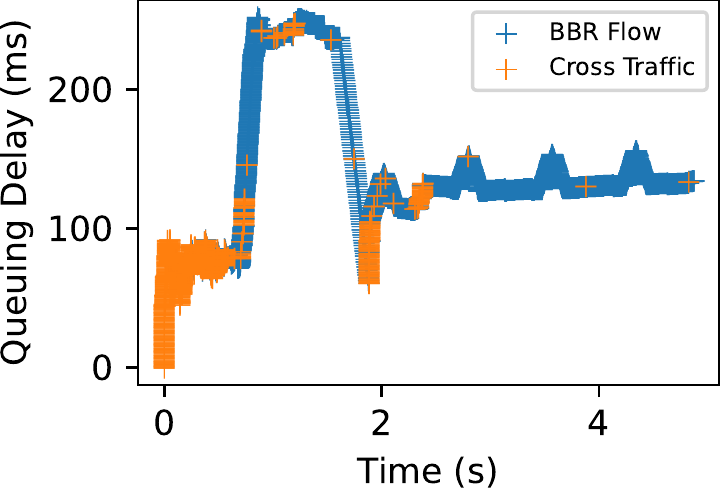}}
    \caption{Analyzing BBR with \ccfuzz.}
    \label{fig:bbr-analysis}
\end{figure*}
We tested NS3's version of TCP-BBR with \ccfuzz, and after a few generations, it produced traces that triggered low throughput for BBR where it get's stuck permanently.
One such trace is shown in Figure~\ref{fig:bbr-traffic}.
For understanding the root cause, we dug into NS3 code and generated various internal logs from BBR's code and from the NS3 TCP socket code.

BBR uses an 8-RTT gain cycle for estimating bandwidth, where it sends at 1.25X the current bandwidth estimate for the first RTT, 0.75X on the second RTT and at 1X for the rest of the gain cycle.
Each RTT is considered as a probing round.
The measured rate in each probing round is processed through a windowed max-filter that keeps the estimates from the last 10 rounds of probing.

We found the root cause to be BBR's mechanism for timing it's bandwidth probing cycles in terms of RTT.
For each packet, the TCP send buffer tracks the number of bytes delivered when that packet was sent in the SKB.
At the beginning of a probing round, BBR records the number of bytes delivered so far.
The probe ends when the prior delivered of the packet most recently ACK (i.e. bytes delivered when the ACKed packet was sent) exceeds the bytes delivered at the beginning of the probing round.

Suppose a packet $P(0)$ is transmitted at time $T_0$, and is lost.
Fast retransmit will cause the first retransmission to occur at some time $T_1 > T_0 + RTT$, and an RTO timer will be set for $T_1 + minRTO$.
At $T_1 + minRTO$, $P(0)$ is retransmitted for the second time.
Suppose $P(i)...P(j)$ were the last few packets sent before the second retransmission for $P(0)$, and the SACKs for these have not arrived yet.
After transmitting $P(0)$ for the second time, $P(i)$ will be transmitted again (a spurious retransmission).
Here, the prior delivered for $P(i)$ is updated in the SKB for $P(i)$ to the current bytes delivered.
If the SACK for the original transmission of $P(i)$ arrives right after the second transmission of $P(i)$, BBR will prematurely end the current probe cycle, since the value of prior delivered for $P(i)$ increased when the spurious retransmission was sent, and now likely exceeds threshold at which the current probing round was supposed to end.
This sequence of events is depicted in Figure~\ref{fig:bbr-timeline}.
Thus, BBR's rate sample is now incorrect, as it is using the time and bytes delivered between the ACK for the original packet, and the packet's spurious retransmission, to calculate the rate.
This can result in a low value for the bandwidth sample.
This can repeat for the other packets $P(i+1)...P(j)$ that were in-flight when $P(0)$ was transmitted the second time. 
If this continues for 10 or more packets, the true bandwidth estimates in the bandwidth max-filter expire, and BBR's bandwidth estimate becomes low.
With a very low bandwidth estimate, delayed ACKs can cause a positive feedback loop, causing BBR to send slower and slower, stalling BBR indefinitely.
It is possible that this is the same issue being referenced in~\cite{strange-question-bbr}.


\ccfuzz was able to trigger this behavior with both, link fuzzing and traffic fuzzing.
Figure~\ref{fig:bbr-link} shows a link trace generated by \ccfuzz that triggers the same bug.
The traffic trace generated by \ccfuzz is very easy to understand - \ccfuzz's implicit constraints on traffic traces generate a clean, minimal trace.
On the other hand, despite our trace annealing mechanism significantly smoothing out the bandwidth variations, the link trace is harder to reason about. 
In the future, we plan to implement better heuristics and implicit constraints in order to generate easier to understand link traces that trigger poor behavior.

In order to try and mitigate this behavior, we made BBR trigger a minRTT probe when an RTO occurs - this slows down BBR momentarily which allows BBR to receive the in-flight ACKs, and thus avoid the spurious retransmissions that cause poor RTT-clocking for BBR's bandwidth probes.
Figure~\ref{fig:bbr-patch} plots the average of the top 20 traces with the lowest throughput in each generation. 
Our proposed fix reduces throughput a little bit, but avoids the permanent stalling behavior observed in BBR without the fix.

\subsection{TCP-CUBIC Incorrect CWND Update}
When testing TCP-CUBIC, we discovered a bug in NS3's implementation of CUBIC's window update during slow start.
When a packet is lost, and it's retransmission triggered by fast retransmit is also lost, the CCA goes into slow start after RTO.
The sender performs a second retransmission for the packet, and when the ACK for this is received, there is a large jump in the cumulative ACK.
CUBIC's slow start window-increase function is called with the large number of segments ACKed.
At this point, the CWND must only be increased upto the slow-start threshold.
In NS3, this check is not performed, and the congestion window is increased by a large value - causing CUBIC to send almost 1-RTO (1 second in the case of NS3) worth of pending data, causing catastrophic losses. 
This leads to CUBIC going into slow start again. As of commit \texttt{60e1e403}, this bug is still present in NS3.
This computation is performed correctly in the Linux kernel source code.

\subsection{Other Findings}
For TCP-Reno, \ccfuzz was able to find a traffic trace similar to the well-known low-rate TCP attack~\cite{kuzmanovic2003low} for a single flow, where traffic bursts cause the same packet sequence to get lost after each retransmission, which triggers exponential RTO back-off. This prevented Reno from ever ramping up after the initial slow start phase.
\ccfuzz can also test CCAs for goals other than low throughput, by just changing the performance component of the scoring function. For example, we ran traffic fuzzing on BBR with the goal of inducing high delays by setting the score function to the 10th percentile delay. This caused \ccfuzz to generate a traffic vector that (1) fills up the queue just before BBR starts, so that BBR cannot see the true link RTT, and (2) injects traffic right after BBR's slow start phase to accelerate queue-growth caused by BBR. This is shown in Figure~\ref{fig:bbr-delay}.

\section{Future Work}
\label{sec:discussion}

\begin{figure}
    \centering
    \subcaptionbox{Valid traces.\label{fig:design:realism-valid}}{\includegraphics[width=0.48\columnwidth]{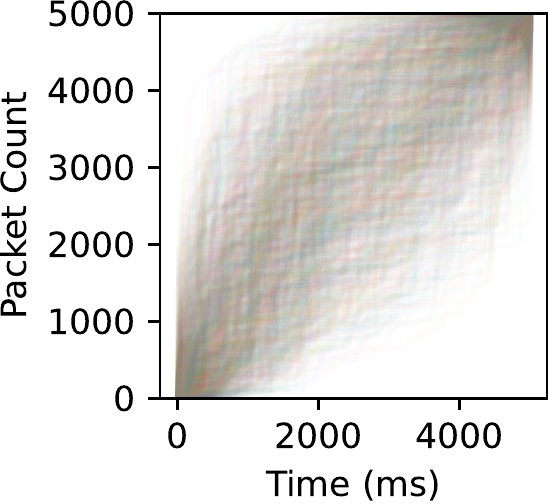}}%
    \hfill\subcaptionbox{Invalid traces.\label{fig:design:realism-invalid}}{\includegraphics[width=0.48\columnwidth]{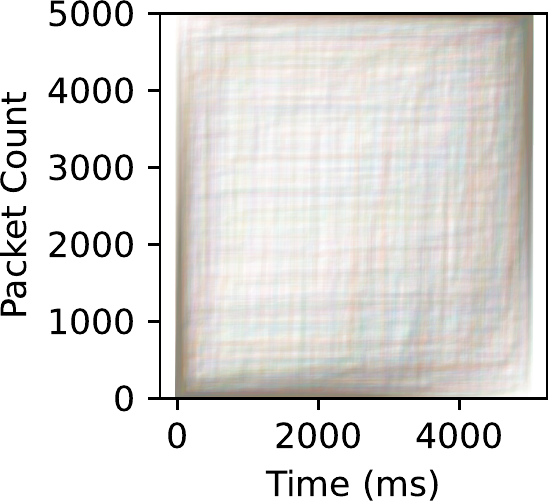}}
    \caption{Distribution of service curves according to realism scores assigned by testing on multiple CCAs. The traces were generated with \textsc{DistPackets}, but without the local rate constraints.}
    \label{fig:design:realism-analysis}
\end{figure}

\textbf{Realism Scoring.}
\label{sec:design:realism-scoring}
The current version of \ccfuzz uses a heuristic-based approach for generating realistic traces.
In the future, we plan to explore an alternate technique for generating realistic traces using aggregate performance across multiple CCAs as a score function to quantify the realism of a trace, assigning high scores to traces under which at least a few algorithms perform well, and vice versa.
Figure~\ref{fig:design:realism-analysis} shows the traces accepted and rejected by this mechanism.
Note how traces that have low bandwidth initially and higher bandwidth later are rejected - such traces will naturally cause low throughput in most CCAs.

In order to reduce the amount of computation required, the realism score can be computed every few iterations instead of every iteration, or can be computed for a single randomly chosen CCA instead of all CCAs in each generation.


\textbf{Diversity and Semantic Scoring.}
Currently, \ccfuzz tends to converge at a point where most traces trigger the easiest to induce performance bug.
In order to find other bugs, an iterative process of fixing the bug and retesting, or defining a score function that negatively weights the manifestation of that bug can be used.
In the future, we plan to explore techniques that automatically result in a diverse set of bugs being found automatically by using machine learning to classify different behavior and dropping traces that trigger similar bugs across generations.
We also plan to implement a framework to translate logical specifications of performance goals into score functions, so that the user does not have to come up with complex score functions themselves in order to make \ccfuzz work.

\textbf{Random Losses and Combined Fuzzing.}
\label{random-losses}
Random packet losses are common on wireless links. \ccfuzz's two modes, link, and traffic fuzzing, do not cover scenarios where random losses occur without a corresponding queue build up.
In the future, we plan to explore loss fuzzing in order to increase the testing coverage, and also explore combining link, traffic, and loss fuzzing into a single process.
Combined fuzzing will result in much more complex network traces that include link variations, cross traffic and loss - these are harder to understand, and thus, it is harder to pin-point the bug.
We plan to address the challenge of generating easier to understand network traces in order to aid debugging.

\section{Conclusion}
In this paper, we have presented the design of an automated congestion control testing tool, \ccfuzz. Our results are highly promising with an initial prototype of \ccfuzz are finding both known and unknown issues with existing well-tested CCAs. We believe that with further development, \ccfuzz could fill an important gap in the development of new CCAs by providing a simple way to identify settings in which a particular CCA performs poorly.

\clearpage

\bibliographystyle{unsrt}
\begin{small}
\bibliography{hotnets22}
\end{small}

\end{document}